\documentclass[12pt]{article}
\pdfoutput=1
\usepackage{graphicx}

\usepackage{amssymb,latexsym}

\def\dalemb#1#2{{\vbox{\hrule height .#2pt
        \hbox{\vrule width.#2pt height#1pt \kern#1pt
                \vrule width.#2pt}
        \hrule height.#2pt}}}

\let\a=\alpha \let\b=\beta \let\g=\gamma \let\d=\delta \let\e=\epsilon
  \let\th=\theta  \let\k=\kappa
\let\l=\lambda \let\m=\mu \let\n=\nu \let\x=\xi  
\let\s=\sigma     
 
        \let\Th=\Theta 
\let\X=\Xi  \let\S=\Sigma  \let\Y=\Psi
 
\let\la=\label  
  
\def\nn{\nonumber} \def\bd{\begin{document}} \def\ed{\end{document}}
\def\ds{\documentstyle} \let\fr=\frac \let\bl=\bigl \let\br=\bigr
\let\Br=\Bigr \let\Bl=\Bigl
\let\bm=\bibitem
\let\na=\nabla
\def\tU{{\widetilde U}}
\let\pa=\partial \let\ov=\overline
\def\ie{{\it i.e.\ }}
\newcommand{\be}{\begin{equation}}
\newcommand{\ee}{\end{equation}}
\def\ba{\begin{array}}
\def\ea{\end{array}}
\def\ft#1#2{{\textstyle{{\scriptstyle #1}\over {\scriptstyle #2}}}}
\def\fft#1#2{{#1 \over #2}}
\def\F#1#2{{ F_{#1}^{(#2)} }}
\def\cF#1#2{{ {\cal F}_{#1}^{(#2)} }}

\def\R{{\bf R}}
\def\sst#1{{\scriptscriptstyle #1}}
\def\oneone{\rlap 1\mkern4mu{\rm l}}
\def\e7{E_{7(+7)}}
\def\td{\tilde}
\def\wtd{\widetilde}
\def\im{{\rm i}}
\def\bog{Bogomol'nyi\ }
\newcommand{\ho}[1]{$\, ^{#1}$}
\newcommand{\hoch}[1]{$\, ^{#1}$}
\newcommand{\bea}{\begin{eqnarray}}
\newcommand{\eea}{\end{eqnarray}}
\newcommand{\ra}{\rightarrow}
\newcommand{\lra}{\longrightarrow}
\newcommand{\Lra}{\Leftrightarrow}
\newcommand{\ap}{\alpha^\prime}
\newcommand{\bp}{\tilde \beta^\prime}
\newcommand{\cB}{{\cal B}}
\newcommand{\cO}{{\cal O}}
\newcommand{\vecx}{\vec{x}}
\newcommand{\vecy}{\vec{y}}
\newcommand{\vecp}{\vec{p}}
\newcommand{\vecq}{\vec{q}}
\newcommand{\tr}{{\rm tr} }
\newcommand{\Tr}{{\rm Tr} }
\newcommand{\NP}{Nucl. Phys. }

\newcommand{\cL}{{\cal L}}
\newcommand{\cA}{{\cal A}}
\newcommand{\cT}{{\cal T}}
\newcommand{\cD}{{\cal D}}

\def\sst#1{{\scriptscriptstyle #1}}
\def\0{{\sst{(0)}}}
\def\1{{\sst{(1)}}}
\def\2{{\sst{(2)}}}
\def\3{{\sst{(3)}}}
\def\4{{\sst{(4)}}}
\def\5{{\sst{(5)}}}
\def\6{{\sst{(6)}}}
\def\7{{\sst{(7)}}}
\def\8{{\sst{(8)}}}
\def\ve{\varepsilon}
\def\vf{\varphi}
\def\F{\Phi}
\def\wg{\wedge}

\newcommand{\tamphys}{\it 
}

\newcommand{\auth}{AUTHORS}

\def\thb{\bar{\theta}}
\def\Thb{\bar{\Theta}}
\def\barp{\bar{p}}
\def\barq{\bar{q}}
\def\barc{\bar{c}}
\def\bard{\bar{d}}
\def\e{\epsilon}

\def \bi{\bibitem}
\def \la {\label}

\def \l {\lambda}
\def\foot{\footnote}
\def \tl  {{\tilde \l}}
\def \sql {{\sqrt \l}}
\def \adss {$AdS_5 \times S^5$\ }
\newcommand{\rf}[1]{(\ref{#1})}
\def \ov {\over}

\def\th{\theta}
\def\Th{\Theta}
\def\vth{\vartheta}
\def\btheta{{\bar\theta}}
\def\ttheta{{{\tilde\theta}}}
\def\bttheta{{{\bar\ttheta}}}
\def\vth{\vartheta}

\def\ra{\rightarrow}
\def\N{{\cal N}}
\def\F{{\cal F}}
\def\uM{\underline{M}}
\def\uA{\underline{A}}
\def\uN{\underline{N}}
\def\uP{\underline{P}}
\def\ua{\underline{a}}
\def\ub{\underline{b}}
\def\uc{\underline{c}}
\def\ud{\underline{d}}
\def\ue{\underline{e}}
\def\uf{\underline{f}}
\def\ui{\underline{i}}
\def\uj{\underline{j}}
\def\uk{\underline{k}}
\def\ual{\underline{\alpha}}
\def\ube{\underline{\beta}}
\def\um{\underline{m}}
\def\un{\underline{n}}
\def\umu{\underline{\mu}}
\def\unu{\underline{\nu}}
\def\ula{\underline{\l}}
\def\uka{\underline{\k}}
\def\usi{\underline{\s}}
\def\urh{\underline{\r}}
\def\cc{\circ}
\def\eqv{\equiv}

\def\ni{\noindent}

\def\Ep{E^{{}^{(+)}}}
\def\Em{E^{{}^{(-)}}}

\def\Mp{M^{{}^{(+)}}}
\def\Mm{M^{{}^{(-)}}}

\def \ha{{1\ov 2}}

\def\r{\rho}

\def\Y{{\rm Y}}
\def\X{{\rm X}}
\def\tY{\tilde{\rm Y}}
\def\tX{\tilde{\rm X}}
\def\dY{\dot{\rm Y}}
\def\dX{\dot{\rm X}}

\def \J {\mathcal{J}}
\def \del {\partial}

\def\dF{\dot{F}}
\def\dG{\dot{G}}
\def\df{\dot{f}}
\def \E {{\cal E}}
\def \S {{\cal S}}
\def \J {{\cal J}}

\def\ms{\mathcal{S}}
\def\mj{\mathcal{J}}
\def\soj{\fr{\ms}{\mj}}
\def \R {{\bf R}}
\def \om {\omega}
\def \bE {\bar E}
\def \x {{\cal X}}

\def \bi{\bibitem}
\def \la {\label}

\def \l {\lambda}
\def\foot{\footnote}
\def \tl  {{\tilde \l}}
\def \sql {{\sqrt \l}}
\def \adss {$AdS_5 \times S^5$\ }
\def \ov {\over}

\def \varpi {{\rm w}}

\def\thb{\bar{\theta}}
\def\Thb{\bar{\Theta}}
\def\mb{\bar{\m}}
\def\ab{\bar{\a}}
\def\zb{\bar{z}}
\def\psib{\bar{\psi}}
\def\barp{\bar{p}}
\def\barq{\bar{q}}
\def\barc{\bar{c}}
\def\bard{\bar{d}}
\def\e{\epsilon}
\def\wb{\bar{w}}
\def\lb{\bar{\l}}
\def\Jb{\bar{J}}
\def\Nb{\bar{N}}
\def\Zb{\bar{Z}}
\def\pab{\bar{\pa}}

\def\At{\tilde{A}}
\def\Bt{\tilde{B}}
\def\Ct{\tilde{C}}
\def\Dt{\tilde{D}}
\def\Et{\tilde{E}}
\def\Ft{\tilde{F}}
\def\Gt{\tilde{G}}
\def\Ht{\tilde{H}}
\def\Mt{\tilde{M}}
\def\at{\tilde{a}}
\def\bt{\tilde{b}}
\def\ct{\tilde{c}}
\def\dt{\tilde{d}}
\def\et{\tilde{e}}
\def\ft{\tilde{f}}
\def\gt{\tilde{g}}
\def\mt{\tilde{\mu}}
\def\nt{\tilde{\nu}}

\def\bA{{\bf A}}

\def\ola{\overleftarrow}
\def\ora{\overrightarrow}
\def\alt{\tilde{\a}}

\def\eh{\hat{e}}
\def\eph{\hat{\e}}
\def\ph{\hat{p}}
\def\alh{\hat{\a}}
\def\beh{\hat{\b}}
\def\gah{\hat{\g}}
\def\muh{\hat{\m}}
\def\nuh{\hat{\n}}
\def\thh{\hat{\th}}
\def\dh{\hat{d}}
\def\ih{\hat{i}}
\def\jh{\hat{j}}
\def\kh{\hat{k}}
\def\deh{\hat{\d}}
\def\wh{\hat{w}}
\def\lah{\hat{\l}}
\def\Ah{\hat{A}}
\def\Ch{\hat{C}}
\def\Omh{\hat{\Omega}}

\def\ps{\rlap{\, /}\;\,p }
\def\ks{\rlap{\, /}\;\,k }

\def\gym{g_{YM}}

\def\adot{\dot{a}}
\def\bdot{\dot{b}}
\def\bpa{\bar{\pa}}

\begin{document}
\overfullrule=0pt
\parskip=2pt
\parindent=12pt
\headheight=0in \headsep=0in \topmargin=0in
\oddsidemargin=0in

\vspace{ -3cm}
\thispagestyle{empty}

 \vspace{0.1cm}

\setcounter{equation}{0}
\setcounter{footnote}{0}
\setcounter{section}{0}

\begin{center}

{\Large\bf Fuzzy Kaluza-Klein induced M2's from a single M5
  }

\vskip 0.8cm

 \vspace{.5cm} { A. J. Nurmagambetov$^{\dagger}$ and I. Y. Park${^*}$\footnote{Permanent address:
  Philander Smith College,
Little Rock, AR 72223, USA
 }  }

\vspace{0.6cm}
{\it A.I. Akhiezer Institute for Theoretical Physics of NSC KIPT\\
1 Akademicheskaya St., Kharkov, UA 61108, Ukraine} ${^{\dagger}}$


\vspace{0.5cm}
{\it Center for Quantum Spacetime, Sogang University\\
Shinsu-dong 1, Mapo-gu, 121-742 South Korea ${^*}$ \\
}

\vspace{0.2cm}
{\it Korea Institute for Advanced Study\\
Seoul 130-012, Korea ${^*}$ \\
}
 \vspace{0.2cm}

{\it APCTP, Postech\\
 Pohang, Gyeongbuk 790-784, Korea ${^*}$}



\end{center}

 \vspace{0.1cm}

 \begin{abstract}
We propose an algorithmic procedure of obtaining multiple M2 brane dynamics starting
with an action of a single M5 brane. The procedure involves a novel Kaluza-Klein
reduction. First, the M5 brane action is truncated to keep
a few leading terms in the derivative expansion. Then
3+3 splitting of dimensions
is carried out.
 With expansion
in terms of the $S^2$ spherical harmonics, the fields are associated with
$SU(N)$ (or its infinite extension) gauge algebra. We present an elaborate
reduction procedure that leads to ABJM theory
when the fuzzy
spherical harmonics are replaced by $SU(N)$ gauge generators.

\end{abstract}
\newpage

\setcounter{equation}{0}
\setcounter{footnote}{0}
\setcounter{section}{0}


\section{Introduction}

 Several years ago, multiple M2 brane actions were proposed
\cite{Bagger:2006sk}\cite{Gustavsson:2007vu}\cite{Aharony:2008ug}.
A system of multiple M2 branes can be polarized into a single M5 brane through
the dielectric-type effect \cite{Emparan:1997rt}\cite{Myers:1999ps}\cite{Constable:1999ac} as shown by the authors
of \cite{Ho:2008nn}\cite{Ho:2008ve}\cite{Ho:2011ni} who
constructed an M5 action starting with BLG action with a ``bottom-up" approach.
We present in this paper what could be viewed as a ``top-down" approach  -- the Kaluza-Klein
induced mechanism
 of obtaining multiple M2 branes starting from a single
M5 brane.\footnote{See \cite{Park:2008qe} and \cite{Bandos:2008fr} for a related discussion. An interesting role played by a Ramond-Ramond C-field 
background was discussed in
\cite{Chu:2010eb} and \cite{Ho:2011yr} in a related context.} We take the M5/M2 case for
an illustration but the procedure should be valid in general.

On an intuitive level, the mechanism is simply to ``cut''
an M5 brane into M2 brane strips.
In this paper we propose that there is a mathematical procedure that
corresponds to the ''cutting``:
the cutting may be realized through a certain regularized discretization of part of the M5
worldvolume as depicted in the figure below.
Roughly, the idea is to compactify the worldvolume theory of a single M5 brane
on a manifold of real dimension three.
The physics of the resulting M2 branes is
correlated with properties of the internal manifold. The choice of the internal manifold determines, among other things, the gauge group of the reduced theory in the case where the harmonics of the internal
manifold can be associated with an infinite extension of certain gauge algebra.

It is well known that the fuzzy spherical harmonics of $S^2$ admit
such infinite extension \cite{hoppe}\cite{de Wit:1988ig}.
As established in those papers, the algebra of the $S^2$ spherical harmonics
can be associated with that of the infinite extension of $SU(N)$.
We choose $S^2\times S^1$
as the internal manifold, and carry out reduction along $S^1$ followed by truncation; we then consider expansion of fields in
terms of the $S^2$ spherical harmonics.
The main reason to choose $S^2$ is for its association with infinite $SU(N)$; the relevance of $S^2$
was discussed in \cite{Nastase:2009ny} and \cite{Gustavsson:2010ep} from different angles.
The worldvolume diffeomorphism turns into the gauge invariance once the fuzzy spherical
harmonics $Y_{lm}$ are manually replaced by $SU(N)$ group generators $\cT^a$.

\begin{figure}[h]
\centering
\includegraphics[width=0.7\textwidth]{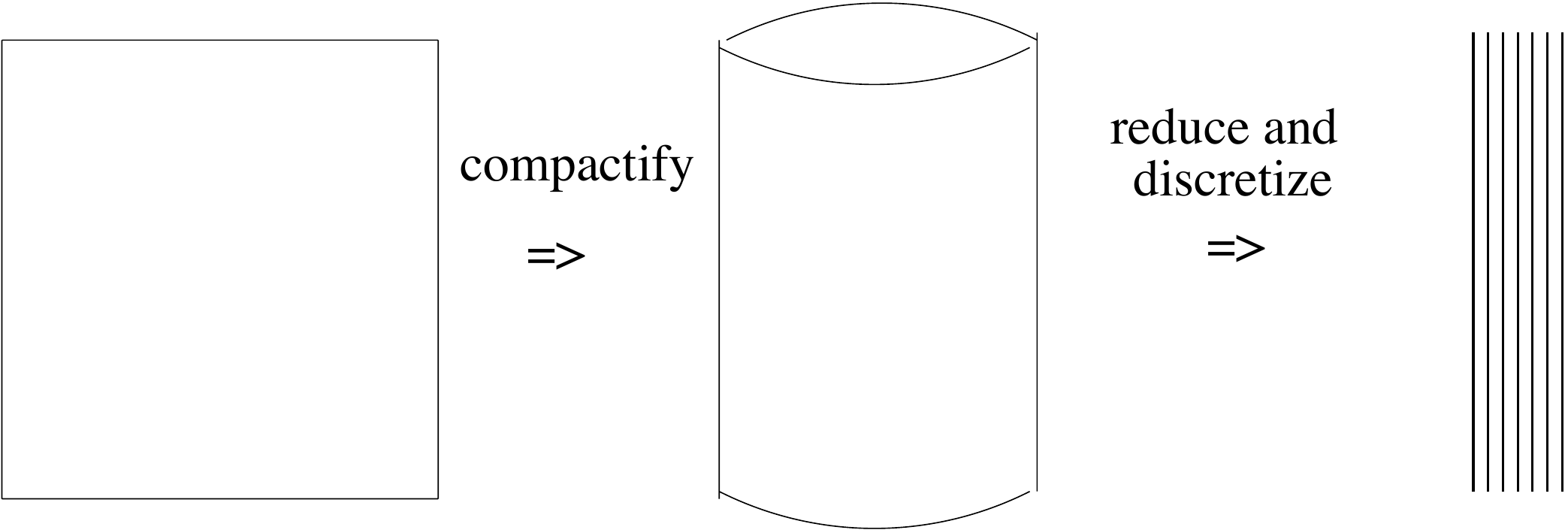}
\caption{From an M5 to M2's} \label{discretization}
\end{figure}


Part of our motivation comes from \cite{Park:2011bg} where certain non-abelianization of the Green-Schwarz open
string action was proposed. It was anticipated that recent progress on the non-abelian M2 brane physics
would shed light on such non-abelianization. We will have more on this in the conclusion.

To some extent, Kaluza-Klein related techniques already appeared in related literature, especially in the bottom-up
approach.
We believe that it is in the top-down approach where the techniques provide
a more enlightening perspective. While the relevance of Kaluza-Klein procedure in
the top-down approach can be seen relatively easily, precise
implementation requires new ideas and components.
Put differently, there are some subtleties in the top-down approach that need to be understood.
For example, an M2 brane theory should have eight
scalars that correspond to the transverse directions of the brane worldvolume. The M5 brane has only
five scalars which correspond to its transverse directions. One may anticipate that the extra scalars would
come, upon reduction, from the self-dual gauge field. On the contrary, the scalars from the self-dual
field should be removed, as we will show. It turns out that
it is certain gauge fixing -- which may be called ``partial static gauge'' -- that brings the deficit scalars.
More specifically, one gauge-fixes only the M2 brane worldvolume coordinates among eleven $X$-coordinates.

Another example of subtleties concerns field ordering. After the M5 brane action
is appropriately reduced, the fields in the action will be Kaluza-Klein expanded in terms of $S^2$ spherical
harmonics, $Y_{lm}$. $Y_{lm}$ can be mapped \cite{hoppe}\cite{de Wit:1988ig} to its
fuzzy version that was denoted therein by $T_{lm}$.
Note that $T_{lm}$'s are functions of $SO(3)$ generators that satisfy the well-known commutation
relations, therefore, {\em non-commuting}. Due to this, one must face ordering issues. As we will discuss, it is likely
that the ordering will be determined uniquely.

The Kaluza-Klein algorithm of this paper offers several new insights on the physics under consideration.
One of them is an alternative understanding of the appearance of the product $U(N)$ group, namely $U(N)\times U(N)$,
in the ABJM model. In the ABJM construction, the product $U(N)$ group arose through a brane construction.
In the current set-up, it appears on a more elementary level: it arises through enforcing reality of the action
after replacement of $Y_{lm}$ (or $T_{lm}$) by $\cT^a$, the group generator.

 Another revealing aspect of the current approach is that it is clearly sets the limitations of the
reduced theory.
 Arriving at the BLG or ABJM type models requires an elaborate reduction
procedure. The fact that such an elaborate procedure is required
reflects that the BLG or ABJM model describes certain aspects of a particular "sector'' of the physical states of M5.
 The starting action of an M5 brane is a Nambu-Goto type that is supposed to describe a single M5 brane
 at its {\em full} energy scale.\footnote{Note, however, that the M5 brane action of \cite{Bandos:1997ui}\cite{Aganagic:1997zq}
needs to be modified by higher derivative corrections to M-theory in ultra-high energy region.}
 As we will explicitly show, cutting out the higher derivative terms is
essential for arriving at BLG or ABJM action (or their variations). This implies that BLG or
ABJM action is only capable of describing the low energy aspects of the M2 brane physics but not the full
energy scale physics \cite{Bagger:2006sk}. (See also \cite{Park:1999xz}.)

\vspace{.3in}
\ni The rest of the paper is organized as follows.
In Section 2, we start with an action for a single M5 brane.
In the derivative expansion, we keep two leading order terms that we name
the DBI term and the $H^2$ term, respectively. Using the technique of \cite{Park:2008qe},
3+3 splitting of dimensions is implemented on the DBI term. After that, 3+3 splitting of the $H^2$ term
is carried out following the steps that could be viewed as
the reverse procedure of \cite{Ho:2008nn}. By generalizing the results of  \cite{Ho:2008nn} to a nonlinear level
and employing certain complexification of the action, we
show in Section 3 that there is a series of reduction ans\"atze that lead to ABJM action.
In the conclusion,
we end with comments on future directions.

\section{Partial reduction of M5 action}

Kaluza-Klein techniques have been employed in the literature, especially in the bottom-up approach.
However, once one gets to the specifics of obtaining multiple M2's from a single M5, certain things
are not so obvious. For example, the single M5 brane action has five scalars after the usual static
gauge-fixing whereas BLG or ABJM action has eight scalars.
One possible way of getting the eight scalars would be the self-dual field's yielding the required scalars
upon reduction. This would be in line with the way in which Kaluza-Klein reduction usually works. It has turned out
that the usual approach does not apply here. The approach that does work is an unusual Kaluza-Klein program
in which a novel gauge-fixing is required. In this section, we deal with this
and a few other related issues, partially carrying out the reduction. The reduction will be completed in the next section.

We start with the worldvolume theory of a single M5 brane formulated in \cite{Bandos:1997ui}\cite{Aganagic:1997zq}.
Several leading terms will be kept in the low energy derivative expansion.
Following \cite{Park:2008qe}\cite{Lee:2010ey},
we consider $(d,n)$ ($d+n=6$) splitting of the dimensions with $d$ denoting
the dimension of the ``external" manifold and $n$ the dimension of the ``internal" manifold.
While doing so, the Nambu bracket naturally appears.
The fact that the Nambu bracket structure is relevant for an M2 brane action is long
known \cite{de Wit:1988ig}\cite{Bergshoeff:1988hw}\cite{Awata:1999dz}. The works of \cite{de Wit:1988ig} and \cite{Bergshoeff:1988hw} were carried out
 in a lightcone framework. It was recently shown in \cite{Park:2008qe} that the Nambu
bracket can be introduced in a more covariant
manner.

In this section, we partially carry out the reduction to $R^{1,2}$ which we take as the external manifold; the
goal is to re-cast the action in (\ref{2apprM5q2}) into a form that is suitable
for the harmonic expansion presented in Section 3.
Let us start with the derivative-expanded form of the covariant action of a single M5 brane \cite{Pasti:1997gx}\cite{Bandos:2000az}
 \bea
S&=&-\int d^6\xi \, \sqrt{-g}\left[ 1-\fr{1}{24}\hat{H}_{mnl}\hat{H}^{mnl}-\fr{1}{8(\pa a)^2}\pa_m a (\hat{H}-\tilde{\hat H})^{mnl}(H-\tilde{\hat H})_{nlp}\pa^p a+\dots \right]\nn\\
&&+\int \Big[ \hat{C}^\6+\fr12 dB^\2\wedge \hat{C}^\3 \Big].
\label{2apprM5q}
\eea
Here $g_{mn}$ is the induced metric on 6D worldvolume, $\hat{H}^\3=dB^\2-\hat{C}^\3$ is the field strength of the worldvolume self-dual gauge field $B^\2$, i.e.
\[
\hat{H}_{mnl}=\tilde{\hat H}_{mnl},\quad \tilde{\hat H}_{mnp}=\frac{1}{3!\sqrt{-g}}\e_{mnprst}\hat{H}^{rst}.
\]
$\hat{C}^\6$ and $\hat{C}^\3$ are the pullbacks of 11D target-space gauge fields.
 One may fix, e.g., the gauge \cite{PST}
\[
\fr{\pa_m a(x)}{\sqrt{-(\pa a)^2}}=\d^0_m .
\]
In this gauge, the action becomes (cf. \cite{Pasti:2009xc})
\bea
S&=&-\int d^6\xi \, \sqrt{-g}\left[ 1-\fr{1}{24}\hat{H}_{mnl}\hat{H}^{mnl}-\fr{1}{8}(\hat{H}-\tilde{\hat H})_0^{~\bar{m}\bar{n}}(\hat{H}-\tilde{\hat H})_{0\bar{m}\bar{n}}+\dots \right]\nn\\
&&+\int \Big[ \hat{C}^\6+\fr12 dB^\2 \wedge \hat{C}^\3 \Big],
\label{2apprM5qfixed}
\eea
where $\bar{m}=1,2,...,5$.
The gauge fixed version of the M5 action \rf{2apprM5qfixed} is equivalent (modulo contribution of
the target-space gauge fields) to that of \cite{Ho:2008nn}
(see \cite{Pasti:2009xc} for details).
We go to flat space and set $\Ch^\6=0,\Ch^\3=0$. Substituting the $a(\xi)$-equation of motion
into \rf{2apprM5q}, the expanded covariant action
reduces to (cf. \cite{Bergshoeff:1996ev})
 \bea
S 
  &=&-\int d^6\xi \,\Big[ \sqrt{-g}-\fr{1}{24}\sqrt{-g}H_{mnl}H^{mnl}+\dots \Big] .
\label{2apprM5q2}
\eea
Now the self-dual field strength is related to its gauge potential in the usual way
\bea
H_{mnl} \equiv (\pa_l B_{mn}+\pa_m B_{nl}+\pa_n B_{lm})
\label{H}\equiv 3\pa_{[m} B_{nl]}.
\eea
The self-duality condition reads
 \bea
 H_{mnp}-\Ht_{mnp}=0,\quad
 \Ht^{mnp}=\frac{1}{3!\sqrt{-g}}\e^{mnprst}H_{rst}
\label{Hsfd6}
\eea
Let us split $m=(\m,i)$,
  \bea
   m=(\m,i)\quad \m=0,1,2 \quad i=1,2,3 \label{is}
  \eea
and denote $\xi^m=(x^\m,y^i)$.
In the next two subsections, we implement 3+3 splittings of the $\sqrt{-g}$-part and the $H^2$-part,
arriving at (\ref{6Dstartpoint2mod2}) as a result.

\subsection{3+3 splitting of $ \sqrt{-\det (g_{mn})}$  part}

The analysis of the Nambu-Goto part, i.e., the first term of (\ref{2apprM5q2}) requires use of two ingredients:
the covariant splitting procedure of \cite{Park:2008qe}\cite{Lee:2010ey} and novel gauge fixing. The gauge fixing procedure
has a certain similarity to that of \cite{Bandos:2008fr}.
Define
 \bea
 S_{NG}&\equiv &-\int d^6 \xi \sqrt{-\det (g_{mn})}=-\int d^6 \xi \sqrt{-\det (\pa_m X^M \pa_n X_M)}  \label{sqrtg}
 \eea
where $M=0,\dots,10$ are the target space indices; $m,n=0,\dots,5$ are the indices on the M5-brane worldvolume.
Let us impose partial gauge fixing
\bea
X^\m=\xi^m \d_m^\m, \quad \m=0,1,2 \label{psg}
\eea
This action can be recast to the form \cite{Park:2008qe}
 \bea
   S_{NG}
 =\int d^3x\, d^3y \;\sqrt{-h}\Big[-h^{\m\n}D_\m X^M D_\n X_M-\fr14 w^{2}\det V+2w
     \Big]  \label{3p3z}
 \eea
where
 \bea
 D_\m X^M \equiv \pa_\m X^M-A_\m^{i} \pa_{i} X^M,\quad i=1,2,3
 \label{D0cov}
 \eea
and the $V$-term is given by
 \bea
\det  V=\fr1{3!}\Big[\e^{i_1,i_2,i_3}\pa_{i_1}X^{M_1}\pa_{i_2}X^{M_2} \pa_{i_3}X^{M_3}\Big]^2 \label{V}
 \eea
The field $w$ is auxiliary, and will play an interesting role later.
Substituting $X^\m=\xi^m \d_m^\m$ explicitly, the equations (\ref{3p3z}), (\ref{D0cov}) and (\ref{V}) become respectively
 \bea
   S_{NG}
 =\int d^3x\, d^3y \;\sqrt{-h}\Big[-h^{\m\n}\eta_{\m\n}-h^{\m\n}D_\m X^{I} D_\n X_{I}-\fr14 w^{2}\det V+2w
     \Big]  \label{Y3p3z2}
 \eea
 \bea
 D_\m X^{I} &\equiv & \pa_\m X^{I}- A_\m^{i} \pa_{i} X^{I}, \quad I=1,...,8
 \label{YD0cov2}
 \eea
 and\footnote{The action \rf{3p3z} implies the following field equation for $A_\m^i$,
 \bea
 {A}_\n^k= (V^{-1})^{ki}(B^T)_{i \n},\qquad V_{ij}=\pa_i X^M \pa_j X_M,~B_{\m i}=\pa_\m X^M \pa_i X_M
 \eea
 The field equations for $(w, h_{\m\n})$ that follow from \rf{3p3z} (or, equivalently, from \rf{Y3p3z2}) are
\bea
w^{-1}=\fr14 \det V, \qquad h_{\m \n}=w^{-1}D_\m X^M D_\n X_M=w^{-1}(\eta_{\m\n}+D_\m X^I D_\n X_I) 
\label{tKmnc}
\eea
Upon substituting these into the lagrangian, one gets the Nambu-Goto form back as can be
easily checked with noting $-\det(g_{mn})=-\det(D_\m X^M D_\n X_M)\cdot \det (V_{ij})$ \cite{Park:2008qe}.
 }
 \bea
\det V=\fr1{3!}\Big[\e^{i_1,i_2,i_3}\pa_{i_1}X^{I_1}\pa_{i_2}X^{I_2} \pa_{i_3}X^{I_3}\Big]^2 \label{YV2}
 \eea
The field strength of $A_{\m i}$ is defined \cite{Park:2008qe} by
  \bea
  \Phi_{\m\n}^i \equiv \pa _\m A_\n^i-\pa _\n A_\m^i
             +(\pa _j A_\m^i) A_\n^j-(\pa _j A_\n^i) A_\m^j
             \label{phimunu}
  \eea
 The nonlinear part of $\Phi_{\m\n}^i$
 will be responsible for the generation of the $A^3$ term of the Chern-Simons action
in the analysis of $H^2$ in the next subsection.
As it stands, the action possesses such rich dynamics - which must be a general aspect of M-brane dynamics - that
the scalar potential $V$ has a spacetime-dependent coupling ''constant``. We
narrow down to a special sector of the M5 dynamics choosing $w=const$. This choice may be taken  as
part of the reduction procedure.

 Let us expand the action \rf{Y3p3z2} over the classical solution to the equations of motion
\bea
h_{\m\n}=w^{-1}(\eta_{\m\n}+D_\m X^I D_\n X^I)
\eea
from which one obtains
\[
h^{\m\n}=w(\eta^{\m\n}-D^\m X^I D^\n X^I+\dots),
\]
and
\[
-h^{\m\n}(\eta_{\m\n}+D_\m X^I D_\n X^I)=-3w+\dots
\]
The action \rf{Y3p3z2} now takes
\[
S_{NG}=\int \, d^3x\, d^3 y~ w^{-3/2}(1-\fr12 D^\m X^I D_\m X^I+\dots)(-w-\fr14 w^2 \det V+\dots)
\]
\[
=\int \, d^3x\, d^3 y~\fr12 w^{-1/2}(\eta^{\m\n}D_{\m}X^I D_{\n}X^I-\fr12 w\det V-2+\dots)
\]
The constant part, -2, can be omitted from the NG part of the action.
 Once $X^M$
are re-scaled according to
 \bea
 X^{I}\rightarrow w^{-\fr14}X^{I} \label{Xscale}
 \eea
the auxiliary field $w$ becomes an overall factor,
\bea
S_{NG}
&=& w_1\int \, d^3x\, d^3 y~ (\fr12\eta^{\m\n}D_{\m}X^I D_{\n}X^I-\fr14 \det V)
\eea
where we have defined
\bea
w_1\equiv \fr1{w}
\eea

\subsection{3+3 splitting of $H^2$ part}

With the analysis in the previous section, the action \rf{2apprM5q2}, now takes the form
 \bea
S &=&w_1\int  \;\Big[\fr12 \eta^{\m\n}D_\m X^I {D}_\n {X}_I-\fr14 \det V-1
     \Big] +\int \Big[H_{mnl}H^{mnl}+\dots \Big]
\label{2apprM5q3}\nn\\
\eea
where the second $\sqrt{-g}$ in \rf{2apprM5q2} has been replaced by one.
(Also the factor $\fr{1}{24}$ has been re-scaled away.)
The reason for this is two-sided. Firstly, we are considering a minimal coupling between
the fields $X$ and $H$. Secondly, it is for the action after the reduction to have two derivatives
at most.  Ultimately, we arrive at
ABJM, and ABJM has up to two derivatives. Put differently, if one keeps $\sqrt{-g}H^2$,
one will get ABJM+higher derivative terms after the reduction
(assuming a consistent procedure still exists).
But then one can drop those higher derivatives terms by going to a low energy limit.

Let us carry out 3+3 splitting of the $H^2$ term. The outcome of the analysis is
given below in \rf{LFe3}. Many of the necessary steps
 can be found in \cite{Ho:2008nn} but the analysis should be run in the reverse, i.e., from a M5 to M2's.
There are two crucial new steps that we have taken.
The first is to identify the field $A_\m^i$ (that has appeared in the ``covariant derivative'' of \rf{D0cov})
with some components of $B_{mn}$. This is in the usual spirit of Kaluza-Klein ans\"atze: the resulting
theory comes to have a reduced solution space, a particular ``sector'' of the original theory.
The second critical step - which is crucial
 for the ``non-abelian'' case - is the adoption of the definition
of $\Phi_{\m\n}$ eq.\rf{phimunu} in the present context.
In \cite{Ho:2008nn}, only the first
two terms of $\Phi_{\m\n}$
appeared because obtaining the quadratic part of
the action was the goal. The last two terms are essential
to produce the $A^3$ terms of the Chern-Simons part (hence the expression ``non-abelian''), as will be
discussed in Section 3.

For convenience, we quote here the definition of the field strength $H$ and the self-duality constraint,
\bea
H_{mnl} \equiv (\pa_l B_{mn}+\pa_m B_{nl}+\pa_n B_{lm}),\quad
\Ht^{mnp}=\frac{1}{3!}\e^{mnprst}H_{rst}
\label{Hq}
\eea
 \bea
 H_{mnp}-\Ht_{mnp}=0,\quad
\label{Hsfd6q}
\eea
 Define
 \bea
 && B_{\m\n}=-\e_{\m\n\r}B^\r,\quad B_{\m j}=A_{\m j},\quad B_{ij}=A_{ij}
 \label{BAna}
 \eea
$A_{ij}$ is a two-form whereas $A_{\m i}$ is a one-form.
Let us take the following reduction ans\"atze that are ``non-abelian'' generalizations of the
corresponding equations of \cite{Ho:2008nn},
\bea
&& H_{\m\n\l}= -\e_{\m\n\l} \pa_\r B^\r,\quad
H_{\m\n i}= \Phi_{\m\n i}-\e_{\m\n\l} \pa_i B_\l \nn\\
&& H_{\m ij}=F_{\m ij},\quad H_{ijk}=F_{ijk}
\label{hphi}
\eea
where
  \bea
  \Phi_{\m\n i} \equiv \pa _\m A_{\n i}-\pa _\n A_{\m i}
             +(\pa _j A_{\m i}) A_{\n j}-(\pa _j A_{\n i}) A_{\m j}
             \label{phimunu_q}
  \eea
and
\[
F_{\m jk}\equiv \pa_{\m}A_{jk}-\pa_j A_{\m k}+\pa_k A_{\m j},\;
  F_{ijk}=(\pa_k A_{ij}+\pa_i A_{jk}+\pa_j A_{ki}), \; A_{ij}\equiv \e_{ijk} A^k
\]

One can re-express the self-duality constraint and the field equation of $H_{mnp}$;
the complete list of the self-duality and field equation of $H$ in terms of $A$'s is as follows.
The self-duality condition $H_{mnk}=\tilde{H}_{mnk}$ yields
 \bea
 \pa^j (F_{\m ij}-\tilde{\Phi}_{\m ij})=0 \label{FmPhit}
 \eea
\[
\pa_\m B^\m = -\fr16 \e^{ijk}F_{ijk},\quad
\Phi_{\m\n i}= \e_{\m\n\l} \pa_i B^\l -\fr12 \e_{\m\n\l} \e_{ijk}F^{\l jk}
\]
Using these, $\pa_m (H^{mnk})=0$ can be put\footnote{
For the moment we consider the free lagrangian $H^2$. What has been achieved by \rf{HAEOM2} is that
the field equations are obtained in terms of fields with the self-duality integrated.
}
\bea
\pa_\m \Ft^{\m\n\l}+\pa_i \Ft^{\n\l i}=0,\quad \pa_\m \Ft^{\m\n k}-\pa_i F^{\n ik}=0,\quad
\pa_\m F^{\m jk}+\pa_i F^{ijk}=0 \label{HAEOM2}
\eea
They should have gauge invariance inherited from the original action (\ref{2apprM5q2}),
the 6D diffeomorphism and $B$-field gauge transformation. Using part of the gauge transformations
(i.e., the part associated with $B_{ij}$ transformation),
let us set
\bea
 A_{ij}=0 \label{Aijgauge}
\eea
 Eq.\rf{HAEOM2} will be supplemented by the
constraint (\ref{FmPhit}).

With the gauge fixing (\ref{Aijgauge}), the first equation of \rf{HAEOM2} becomes $\pa_i \Ft^{\n\l i}=0$.
As can be seen by explicitly writing, the left-hand side vanishes identically.
The third equation of \rf{HAEOM2} simplifies to
 \bea
 \pa_\m F^{\m jk}=0
 \eea
yielding a Lorentz type gauge after the reduction.
 The second equation of  \rf{HAEOM2} can be reproduced by
an action. To see that, let us consider
\bea
\mathcal{L}_{A}&=& w_2 \Big[- F_{\m ij} F^{\m ij}-2 \e^{\m\n\l} \e^{ijk} \pa_\m A_{\n i} \pa_j A_{\l k}
+\fr{1}2 \e^{\m\n\l} \e^{ijk} F_{\m ij}\Phi_{\n\l k}\nn\\
&& + \e^{\m\n\l} \e^{ijk} A_{\m j}\pa_k  (A_\l^s \pa_s A_{\n i}-A_\n^s \pa_s A_{\l i}) \Big]
\label{LFe3b}
\eea
and compute $A_\m^i$ variation. $w_2$ is a numerical constant to
be determined later. This action should be supplemented by the constraint \rf{FmPhit}.
The $X$-part of the action contains $A_\m^i$ inside the covariant derivative \rf{D0cov}. Here we
focus only on the gauge part of the action, $\cL_A$. An advantage of \rf{LFe3b} over the original $H^2$ form
is that the self-duality condition can be explicitly implemented in the \rf{LFe3b} as we will see below.
Also by going from $B_{\m i}$ base to $A_{\m i}$ base, the gauge field equation of the coupled
system of \rf{2apprM5q3} is obtained by $A_{\m i}$-variation as a fully legitimate procedure.

Variation of the first three terms yield, respectively,
\bea
\d(- F_{\m ij} F^{\m ij})=-4 \d A_{\m j}\pa_i F^{\m ij} \label{1stterm}
\eea
\bea
&& \d \left(-2 \e^{\m\n\l}\e^{ijk}\pa_\m A_{\n i}\pa_j A_{\l k}\right)=\e^{\m\n\l}
                  \e^{ijk}\d A_{\m j} \pa_\n F_{\l ki}
+ \e^{\m\n\l} \e^{ijk} \d A_{\m j}\pa_k \Phi_{\n\l i} \nn\\
&&\hspace{2in}- \e^{\m\n\l}
\e^{ijk} \d A_{\m j}\pa_k (A_\l^s \pa_s A_{\n i}-A_\n^s \pa_s A_{\l i})
\label{2ndterm}
\eea
\bea
&& \d \left( \fr{1}2 F_{\m ij} \e^{\m\n\l} \e^{ijk}\Phi_{\n\l k} \right)
=- \d A_{\m j} \e^{\m\n\l} \e^{ijk} \pa_k \Phi_{\n\l i} - A_{\m j}
     \e^{\m\n\l} \e^{ijk} \pa_k \d \Phi_{\n\l i} \label{3rdterm}
\eea
Noting that
\bea
&& - A_{\m j} \e^{\m\n\l} \e^{ijk} \pa_k \d \Phi_{\n\l i}\nn\\
=&&  \d A_{\m j} \e^{\m\n\l} \e^{ijk} \pa_\n F_{\l ki}-  A_{\m j}
     \e^{\m\n\l} \e^{ijk} \pa_k \d (A_\l^s \pa_s A_{\n i}-A_\n^s \pa_s A_{\l i})
\eea
one gets
\bea
&& \d \left( \fr{1}2 F_{\m ij} \e^{\m\n\l} \e^{ijk}\Phi_{\n\l k} \right)
=- \d A_{\m j} \e^{\m\n\l} \e^{ijk} \pa_k \Phi_{\n\l i}
      + \d A_{\m j} \e^{\m\n\l} \e^{ijk} \pa_\n F_{\l ki}\nn\\
 &&  \hspace{2in} -  A_{\m j} \e^{\m\n\l} \e^{ijk} \pa_k \d (A_\l^s \pa_s A_{\n i}-A_\n^s \pa_s A_{\l i})
 \label{2ndterm2}
\eea
Variation of the last term of \rf{LFe3b} cancels against terms in \rf{2ndterm2} and \rf{2ndterm}.
Gathering the results above, one gets
\bea
\d \cL_A= 4w_2\, \d A_{\m j}\left(\pa_\n \Ft^{\n\m j}-\pa_i F^{\m ij}\right)
\eea
This completes the proof of the claim.
The action given in \rf{LFe3b} can be re-expressed as
\bea
\mathcal{L}_{A}&=& w_2 \Big[- F_{\m ij}\left( F^{\m ij}
       -\tilde\Phi^{\m ij}\right)-2\e^{\m\n\l} \e^{ijk} \pa_\m A_{\n i} \pa_j A_{\l k} \nn\\
&& + \e^{\m\n\l} \e^{ijk} A_{\m j}\pa_k  (A_\l^s \pa_s A_{\n i}-A_\n^s \pa_s A_{\l i}) \Big]
\label{LFe3a}
\eea
or, on account of \rf{FmPhit},
\bea
\mathcal{L}_{A}&=& w_2\Big[-2 \e^{\m\n\l} \e^{ijk} \pa_\m A_{\n i} \pa_j A_{\l k}+2 \e^{\m\n\l} \e^{ijk} A_{\m j}
   \pa_k  (A_\l^s \pa_s A_{\n i} )\Big]
\label{LFe3}
\eea

\section{Fuzzy Kaluza-Klein compactification   }

 The analysis of the previous section puts the action (\ref{2apprM5q3}) in the form
 \bea
 && \int d^3 x d^3 y\; \left\{
 w_1\Big[\fr12 D_{\mu}X^I D^{\mu}X_I-\fr{1}4 \Big(\e^{i_1,i_2,i_3}\pa_{i_1}X^{I_1}\pa_{i_2}X^{I_2} \pa_{i_3}X^{I_3}\Big)^2
    -1 \Big] \right. \nn\\
 && \left. +
 w_2\Big[-2 \e^{\m\n\l} \e^{ijk} \pa_\m A_{\n i} \pa_j A_{\l k}
      +2 \e^{\m\n\l} \e^{ijk} A_{\m j}\pa_k  A_\l^s \pa_s A_{\n i}  \Big] \right\}
             \label{6Dstartpoint2mod2}
 \eea
The action \rf{6Dstartpoint2mod2} is a {\em partially} reduced form in the sense that the fields still
depend on all 6D coordinates. The worldvolume is taken as $R^{1,2}\times (S^2\times S^1)$.
The reduction will be completed in this section and it will be shown
that a mechanism or an algorithm exists whereby the action (\ref{6Dstartpoint2mod2}) can be converted
into a BLG or an ABJM-type action.

\subsection{generalities}

One of the centermost parts of the remaining reduction procedure
is expansion in terms of the harmonic functions of the internal manifold.
Before we get to the details of the expansion, let us address several
issues that merit discussion.
In Kaluza-Klein compactification,
 one considers expansion in the modes of the internal manifold,
 \bea
 \phi(x,y)=\sum \phi^a(x)H_a(y) \label{KK}
 \eea
where $\phi$ is a generic field in (\ref{6Dstartpoint2mod2}), and $H_a(y)$
is a schematic notation for the spherical harmonics.
Necessarily, the physics of the M2 branes will be correlated with the properties of the internal manifold, and
the final form of the M2 action depends on the choice of the internal manifold.
Then a question arises with regard to the internal manifold to choose.
The choice largely determines the physics that the reduced action is to describe.
Reversely, if one has a certain M2 brane physics to describe, the internal manifold must be chosen accordingly.

Let us illustrate the point with the current case.
 Suppose that we intend the M2 brane system to have a gauge group of $SU(N)$
nature (such as $SU(N)\times SU(N)$) or its infinite extension. Given the relation between $S^2$ and $SU(\infty)$
\cite{de Wit:1988ig}, it is not difficult to infer that the internal manifold should involve $S^2$: we
are naturally led to some type of a fibration over $S^2$.
 The simplest fibration is $S^2\times S^1$.\footnote{
One may be concerned that the involvement of $S^1$ - which is one of the worldvolume directions
of the M5 brane - may imply that the resulting non-abelian action may be that of D-branes instead
of fundamental M-branes. This will not be the case. To see that, one may start with an NS5 brane action,
and repeat the analysis of the present paper; most of the analysis would carry over with only
minor modifications. Since an NS5 brane is not related to a D-brane via T-duality, the nonabelian
action should be that of fundamental branes such as M2's or fundamental strings upon further reduction.
}
This choice was also influenced by the work of \cite{Gustavsson:2011mg}. A more complicated choice
such as (Hopf-fibrated) $S^3$ may be possible. It is just that the resulting M2 brane physics would
 be more exotic. (For example, unlike $S^2$, $S^3$ is not associated with an
infinite extension of any
Lie algebra to our knowledge.)

Our proposal is to replace the harmonics $Y_{lm}$ by its fuzzy counterpart $T_{lm}$, and subsequently
 $T_{lm}$ by the $SU(N)$ gauge generator $\cT^a$. The replacement would lead to ordering
ambiguities in the action since $T_{lm}$'s do not commute. The ambiguity
is not present for the quadratic terms because $\int d^3y$ will be replaced by $\Tr$, the trace over the gauge generators.
 We require (and expect) as part of our proposal that the ambiguity should be (mostly) resolved
by symmetries - such as gauge symmetry and supersymmetry - that are expected for an M2-brane system.
They, with the index structures of the fields, will greatly reduce the number of
possibilities. We believe that they will
 determine the final theory uniquely, at least, for the current case: it seems implausible
to have two different theories that share the kinetic terms and the symmetries but differ 
in the orders of the fields that appear in the
interactions.

\subsection{complexification leading to ABJM type theories}

We now turn to specific cases:
reduction to ABJM type actions. First, we take a reduction procedure that leads to ABJM theory.
Evidently there are some manipulations required to proceed
from (\ref{6Dstartpoint2mod2}) to ABJM action.
 There exist two main differences between (\ref{6Dstartpoint2mod2}) and
ABJM action. The first is that the action
of (\ref{6Dstartpoint2mod2}) is in terms of real fields whereas ABJM action is in terms of
complex fields. This implies that some kind of
complexification is required.
(Indeed, complexification is a very important step; it will bring the product $U(N)$ group as will be discussed shortly.)
The second difference lies in the gauge symmetries.
The action (\ref{6Dstartpoint2mod2}) has gauge symmetry that is inherited from the original
action (\ref{2apprM5q2}). (Part of the gauge symmetry has been used for partial gauge
fixing.) Therefore, what seems
required is replacement of $Y_{lm}$ by the group generator $\cT^a$.
In fact, the first and the second requirements feed off of each other.
(The action (\ref{6Dstartpoint2mod2}) differs from ABJM in that it has derivatives along the internal
directions. This difference is related to the second difference, and will be dealt with by further reduction.)

The real action (\ref{6Dstartpoint2mod2}) can be expanded in terms of the spherical harmonics which are complex.
The replacement is not without the subtlety that is associated with the reality of the action. When the action is ``Fourier" expanded,
the reality is assured
by imposing certain constraints on the ``Fourier coefficients". However, the same constraints cannot
work after replacement by the group generators. To remedy the problem, we propose adding complex conjugate
terms explicitly to the action, viewing the fields in (\ref{6Dstartpoint2mod2})
complex.
With the proposed ``complexification", the fields in (\ref{6Dstartpoint2mod2}) can be put into the form of \rf{KK}.

 Since the matrix valued fields in ABJM action do not commute, the regular spherical
harmonics $Y_{lm}$ should be made non-commuting somehow. Given the work of \cite{hoppe}\cite{de Wit:1988ig}, the
natural step for this would be to replace $Y_{lm}$
by its fuzzy version, $T_{lm}$, and then to replace $T_{lm}$ by the gauge group generator $\cT^a$ afterwards.
The use of $T_{lm}$ means that the two-sphere $S^2$ is actually fuzzy.
We will discuss ordering ambiguities as they arise.

\vspace{.3in}

There is a relatively simple complexification procedure.
In this procedure, we replace the gauge field
according to
\bea
 A_\m^i \rightarrow -i\cA_\m^i+i\hat{\cA}_\m^i
  \label{AtocA}
\eea
and treat $\cA_\m^i$ and $\hat{\cA}_\m^i$ independently.
Let us consider complexification of
the scalar part first and the gauge part afterwards.
Due to $SO(8)$ triality, one can regard $X^I$ in (\ref{6Dstartpoint2mod2})
as complex. (See the discussion towards the end for more details.)
The next step is to replace $Y_{lm}$ with $T_{lm}$. As soon as one
considers the replacement, one faces a few subtleties, some of which have been pointed out above.
One subtlety is that not only does the ``bare" $Y_{lm}$ appear but so does $Y_{lm}$ with
derivatives along the internal directions. Recalling the
works of \cite{hoppe} and \cite{de Wit:1988ig} and utilizing some of the results of \cite{Gustavsson:2011mg}
provide a hint as to what should be done to the derivatives.
 The authors of \cite{hoppe} and \cite{de Wit:1988ig} considered the fuzzy version
 of $Y_{lm}$ denoting it $T_{lm}$, and defined the
corresponding structure constant by considering $[T_{lm}, T_{l'm'}]$ with $l\leq N-1$. The structure constant defined
this way approaches the structure constant
of the Nambu bracket $\{Y_{lm}, Y_{l'm'}\}$ when $N\rightarrow \infty$.
The result naturally suggests that the Nambu bracket of $Y_{lm}$'s should be replaced by a commutator
of the gauge generator.

Replacement of $Y_{lm}$ by $T_{lm}$ is an intermediate step: $T_{lm}$ should be replaced by the gauge
generator $\cT^a$. More specifically, we propose
 \bea
X_{lm}^{I} T^{lm} &\rightarrow & X_a^I (\cT^a)_{\b\g}\rightarrow X^I_{\b\g}
 \eea
where the adjoint generator $\cT^a$ is that of the first factor of $SU(N)\times SU(N)$; there are analogous relations
for the complex conjugate fields with the gauge generators belonging to the second $SU(N)$.
Note that with the intermediate step $ X_a^I (\cT^a)_{\b\g}$, the $(\b,\g)$ indices are associated with
single $SU(N)$.  However, the redefined field $X^I_{\b\g}$ may be viewed as a bi-fundamental representation
of $SU(N)\times SU(N)$ with each of $(\b,\g)$ corresponding to each factor of $SU(N)$.

The proposed reduction for the $V$-term can be based on the result just stated. The $V$-term is of the three-bracket form with internal
derivatives acting on $X^I$'s. Assuming $S^2\times S^1$ as the internal manifold, one can
show \cite{Gustavsson:2011mg} that the three-bracket of the potential $V$ can be re-expressed in terms
of the two-bracket.
Based on this, we replace $Y_{mn}$ (or its fuzzy version $T_{mn}$ ) by the gauge generator $\cT^a$,
 \bea
  \Big(\e^{i_1,i_2,i_3}\pa_{i_1}X^{I_1}\pa_{i_2}X^{I_2} \pa_{i_3}X^{I_3}\Big)^2
  \rightarrow  \Tr ([X^{I_1},X^{I_2},X^{I_3}][\bar{X}_{I_1},\bar{X}_{I_2},\bar{X}_{I_3}])
\label{Vinga}
 \eea
with an appropriate numerical constant in front. This is the same as the potential term of ABJM.
Therefore, one arrives at a lagrangian that is the same as the corresponding part of
ABJM theory.

Now let us turn to the gauge part.
What directly came out of the Kaluza-Klein analysis in Section 2 was the real-field based lagrangian.
 Let us view $A_\m^i$ as a complex field
 renamed $\cA_\m^i$, and explicitly add a complex conjugate part to enforce the reality:
 \bea
 {\cL}_{Ac}&=& -2w_2 \e^{\m\n\l} \e^{ijk} \pa_\m \cA_{\n i} \pa_j \cA_{\l k}
      +2w_2 \e^{\m\n\l} \e^{ijk} \cA_{\m j}\pa_k  (\cA_\l^s \pa_s \cA_{\n i} ) \nn\\
&& \hspace{4.4cm}
+c.c.
 \label{LFe3c}
 \eea
now,
\bea
{\cal F}_{\m ij}\equiv -\pa_i \cA_{\m j}+\pa_j \cA_{\m i}
\eea
As a matter of fact, the replacement \rf{AtocA} in the gauge part of the action produces cross terms.
Since the integration $\int d^3 y$ will be replaced by trace over the product gauge group eventually, those
terms will contain trace over a single gauge generator, therefore vanish. For this reason, they have
 been omitted from \rf{LFe3c}.

As mentioned above, the ordering ambiguities should be resolved based on the
anticipated symmetries. More specifically, for the quadratic terms in (\ref{6Dstartpoint2mod2}), there is
no ambiguity since there is trace over the internal space.  For the $\cA$-cubic terms, the ordering causes
overall sign ambiguities. It should be possible to fix the sign based on various symmetries such as gauge
symmetry and/or supersymmetry.
For the sextic potential, again, it would be an overall sign issue which would be resolved in a manner similar to
that of the cubic terms.

\ni For further reduction, consider
 \bea
 \cA_\m^i(x^\m,y^j) = \cA_\m(x^\m,y^{\jh})\; e^{i\b_i y^3},\quad \jh=1,2
 \eea
where $\b_i$'s are constants to be determined. With this ansatz, one can show that the quadratic term reduces
 \bea
-2 \e^{\m\n\l}\e^{ijk}\pa_\m \cA_{\n i}\pa_j \cA_{\l k}
 & = & -2i(\b_1-\b_2)e^{i(\b_1+\b_2)y^3} \,\e^{\m\n\l}  \Big[ \pa_\m\cA_{\n } \cA_{\l } \Big]
\label{CS1}
\eea
Let us take
 \bea
 \b_1+\b_2=0
 \eea
The cubic term requires more lengthy, although straightforward, algebra; one can show
\bea
&& 2 \e^{\m\n\l} \e^{ijk} \cA_{\m j}\pa_k  (\cA_\l^s \pa_s \cA_{\n i})\nn\\
 =&&
  \;\; 2i\Big[\b_1 e^{i\b_1y^3}-(\b_1+\b_3)e^{-i(\b_1-2\b_3)y^3}\Big]\e^{\m\n\l} \cA_\m \cA_\l \pa_1\cA_\n  \nn\\
 &&
  +2i\Big[\b_1 e^{-i\b_1y^3}-(\b_1-\b_3)e^{i(\b_1+2\b_3)y^3}\Big] \e^{\m\n\l} \cA_\m \cA_\l \pa_2\cA_\n \nn\\
  && +\;\; 2\b_1\b_3 e^{i\b_3 y^3}\, \e^{\m\n\l} \cA_\m\; \cA_\n \cA_\l
\label{rct}
\eea

By making, e.g., the following choices
 \bea
 \b_1=1=-\b_2,\quad \b_3=\fr32
 \eea
the first two terms in \rf{rct} are removed by $\int_0^{2\pi} dy^3$ whereas the third term becomes
\bea
4i w_2 \e^{\m\n\l}  \cA_\m\; \cA_\n \cA_\l .
\eea
Putting the results of the quadratic term and the cubic term together, the action \rf{LFe3c}
takes, after reduction,
\bea
 {\cal L}_{Ac} =-4 i w_2 \,\e^{\m\n\l} \left(2\pi \pa_\m\cA_{\n } \cA_{\l }
    -\cA_\m\; \cA_\n \cA_\l \right)+c.c. \label{abjmCSp}
\eea
Each field $\cA^\m$ above can be expanded in terms of the fuzzy spherical harmonics,
\bea
\cA_\m= \cA_\m^{lm}T_{lm}
\eea
The final step is to replace $\cA_\m^{lm}T_{lm}$ by $\cA_\m^a \cT^a$
 \bea
 \cA_{lm}^\m T_{lm}&\rightarrow & \cA_{a}^\m \cT^a
 \eea
 arriving at the corresponding part of
ABJM action up to the numerical coefficients. The numerical
coefficients can be adjusted by noting that \rf{abjmCSp} can be rewritten as
\bea
 {\cal L}_{Ac} =-16\pi i w_2 \,\e^{\m\n\l} \left(\fr12 \pa_\m\cA_{\n } \cA_{\l }
    -\fr{g}{4\pi} \cA_\m\; \cA_\n \cA_\l \right)+c.c.
\eea
where $g$ is a constant. This is possible by making a re-scaled reduction ansatz in
Section 2.2. Therefore, it does not affect $\cA$ field in the the covariant derivative.
Now set
\bea
g= -\fr{4\pi i}{3},\quad  w_1=k,\quad w_2=i\fr{k}{16 \pi }
\eea
where $k$ is the Chern-Simons level parameter.

Finally, we briefly note that there is a slightly different complexification that one can take.
The final outcome should be equivalent to ABJM theory by trivial gauge field redefinition.
Starting with \rf{LFe3c}, the scalar part may be done differently with the gauge part of the analysis unmodified.
One can derive a complex version of the scalar part (\ref{6Dstartpoint2mod2}) using some of the
ingredients in \cite{Gustavsson:2011mg}. (The {\em complexified} version will then be the resulting complex action plus
its complex conjugate.)

Let us introduce
 \bea
 Y^M=\left(
\begin{array}{c}
Y^\m  \\
Y^{\a}
\end{array}
\right)  \quad \mbox{with} \quad Y^\m=X^\m, \quad Y^\a=\left(
\begin{array}{c}
Z^A \,e^{i\s}\\
Z_A \,e^{-i\s}
\end{array}
\right)  \label{xz}
 \eea
  where $\a$ is a $SO(8)$ spinor index and $A$ is a $SU(4)$ index.
Due to the $SO(8)$ triality, \rf{sqrtg} can be re-expressed \cite{Gustavsson:2011mg}
as\footnote{This can be seen by the definition \rf{xz} and
using $Y_\a=(Y^\a)^*$ which was stated in \cite{Gustavsson:2010ep}.}
 \bea
 S_{NG}&\equiv &-\int d^6 \xi \sqrt{-\det (\pa_m Y^M \pa_n Y_M)} \label{Yaction}
 \eea
Following the steps that are analogous to those of the real case before, one gets
 \bea
   S_{NG}
 =w^{-1}\int d^3x\, d^3y \;\Big[\eta^{\m\n}D_\m Z^A D_\n \Zb_A-\fr14 \det V-1
     \Big]  \label{3p3resc}
 \eea
where
 \bea
 D_\m Z^A \equiv \pa_\m Z^A-\cA_\m^{i} \pa_{i} Z^A
 \label{ZD0cov}
 \eea
and
 \bea
V=\fr1{3!}\Big[\e^{i_1,i_2,i_3}\pa_{i_1}Z^{A_1}\pa_{i_2}Z^{A_2} \pa_{i_3}Z^{A_3}\Big]^2 \label{ZV}
 \eea

\section{Conclusion}

So far in the literature, the popular approach (see e.g., \cite{Ho:2008nn}) of relating
an action of a single M5 and that of multiple M2 branes has been through the Myers' type
effect.\footnote{The only exception that we are aware of is \cite{Bandos:2008fr} where
the starting point is an M5 action. }
In that approach, one builds up an action of a single M5 brane starting from a non-abelian
M2 brane action. The Kaluza-Klein procedure of this paper - which can be viewed
as the reverse procedure of a generalized Myers' effect - offers an alternative perspective.

What we have established is a mathematical procedure that corresponds to
``cutting" the M5 brane into pieces with each piece being an M2 brane.
The mathematical procedure is to compactify part of the M5 brane
worldvolume and discretize it by introducing ``non-commutativity" in the compactified worldvolume.
To this end, we have started with the covariant M5 action constructed by \cite{Bandos:1997ui}\cite{Aganagic:1997zq}, and have
carried an elaborate Kaluza-Klein reduction taking $S^2\times S^1$ as an internal manifold. To
make a connection with (infinite) $SU(N)$ algebra, the $S^2$ part has been taken fuzzy.
The fact
that it takes an elaborate reduction procedure should reflect that the resulting
theory describes a narrow and special sector of the M2 brane dynamics, even more so the original M5 brane dynamics.

The choice of $S^2\times S^1$ was for simplicity.
One may consider an internal manifold with different $S^1$ fibering over $S^2$.
In general, a different fibering would lead to a different theory once the $S^1$ direction is reduced.
In the case where only the {\em lowest} Kaluza-Klein mode
is kept (i.e., reduction followed by truncation), there might be a corresponding reduction
for each fibering  that yields the same reduced theory. Settling this issue completely will require more work.

Let us pause to contemplate a potentially interesting issue.
 As we have shown in this letter, ABJM theory can be constructed by reducing a single M5 action
on $S^2\times S^1$.
There are characteristics of the {\em reduced theory}, i.e., ABJM theory, that provide hints as to its origin.
 It has a fuzzy $S^2$ solution that displays features
of $S^3$ (that can be constructed through Hopf fibration over $S^2$) as against the simpler $S^2 \times S^1$.
With regard to this difference (i.e., $S^3$ vs $S^2\times S^1$), there seem to be two possibilities.
The first possibility is that one should actually consider $S^3$ reduction of a single M5; there may be
a different, more complicated reduction procedure that leads to ABJM theory.
The second possibility is that the appearance of $S^1$ Hopf fibration over $S^2$ may be associated with
one of the limitations of ABJM theory.
In other words, the appearance of Hopf fibration may be a peculiarity due potentially to the incapability
of the reduced theory to describe the full physics of the original theory. 
Given the profound relation between gauge degrees of freedom and
gravity degrees of freedom, study of 11D supergravity
solutions would be useful to settle this issue.
Stable $S^3$ compactification of 7D gauged N=2 Supergravity does not exists (see e.g. \cite{Pernici:1984nw})
while 6D gauged N=2 Supergravity admits stable  $S^2$ compactification \cite{Salam:1984cj}. 
Since those 7D- and 6D- supergravities are related by $S^1$ reduction, it follows that the 7D supergravity
 admits the stable compactification on $S^2 \times S^1$. This seems consistent with our result, and to point towards the second possibility above.

As briefly mentioned in the introduction, part of our motivation comes from \cite{Park:2011bg} where certain
non-abelianization of the Green-Schwarz open string action was proposed. Some of the ingredients of the current
work will be useful for providing foundations for (or even first-principle derivation of) the proposal.
To that end, two things must be done. For the current work, we kept only a few leading terms
in the derivative expansion. That step must not be taken in order to obtain a nonabelian string action starting from an M5.
Secondly, the currently work should be generalized to a curved background.
It is one of the near-future directions that we are taking.

\vspace{1in}

\ni {\bf Acknowledgements}\\
Work of AN is supported in part by the Joint DFFD-RFBR Grant \# F40.2/040.
He thanks I. Bandos and D. Sorokin for valuable discussions.
IP thanks the hospitality of the members of CQUeST during his stay.
This work greatly benefited from
discussions with J.H. Park. He also thank P.M. Ho for his hospitality during the visit to
National Center for Theoretical Sciences, Taipei, Taiwan.

\newpage


\end{document}